\documentstyle[psfig]{laa}
\newcommand{\bildh}[3]{{\psfig{figure=#1,height=#2,#3}}\label{#1}}

\begin{document}
\thesaurus{01(03.20.4; 08.09.2 Eta Carinae; 08.03.4; 08.09.1)}
\title{Speckle-masking imaging polarimetry of $\eta$ Carinae: 
evidence for an equatorial disk}
\author{Heino Falcke\inst{1,2}, Kris Davidson\inst{3}, Karl-Heinz Hofmann\inst{1}, Gerd Weigelt\inst{1}}
\institute{Max-Planck Institut f\"ur Radioastronomie, Auf dem H\"ugel 69, 
D-53121 Bonn, Germany (weigelt@mpifr-bonn.mpg.de)
\and
Department of Astronomy, University of Maryland, College Park, MD
20742-2421, USA (hfalcke@astro.umd.edu)
\and
Astronomy Department, University of Minnesota,
116 Church St., Minneapolis, MN 55455, USA  }

\date{Accepted for publication in A\&A Letters (Dec. 19, 1995) -- in press}

\maketitle
\markboth{Falcke et al.: Speckle-masking imaging polarimetry of $\eta$ Carinae}{Falcke et al.: Speckle-masking imaging polarimetry of $\eta$ Carinae}

\begin{abstract}
With our new speckle imaging polarimeter we have obtained the first
polarimetric images with sub-arcsecond resolution of the Luminous Blue
Variable $\eta$ Carinae in the H$\alpha$ line. The polarization
patterns at the 3'' scale match well earlier conventional imaging
photometry and can be interpreted as Mie scattering. In 
crosscorrelation-centered
images we detected in polarized light a bar in the
NE part of the equatorial plane of $\eta$ Carinae. High-resolution
0.11'' polarimetric speckle reconstructions reveal a compact structure
elongated in the same direction which is consistent, in degree and
position angle of the polarisation, with the presence of a
circumstellar, equatorial disk. The degree of polarization of the
previously discovered speckle objects and the H$\alpha$ arm is
relatively low ($\sim10\%$) and thus may indicate a position within
the equatorial plane. We also discovered a highly polarized
($20\%-40\%$) bipolar structure along the major axis of the Homunculus
nebula which can be traced down to the sub-arcsecond scale. This is
probably the inner part of a bipolar outflow into the Homunculus.

\end{abstract}

\section{Introduction}
Because of its extraordinary luminosity of $\sim10^{6.6}L_{\odot}$ the
 Luminous Blue Variable $\eta$ Carinae is one of the most interesting
 objects for the understanding of the late evolutionary stages of
 massive stars (see Humphreys and Davidson 1994, and references
 therein). It is embedded in the Homunculus, a bi-polar nebula
 oriented at PA 132$^\circ$, which is reflecting light from the
 central object. $\eta$ Carinae was also one of the earliest complex
 structures that was successfully studied at sub-arcsec resolution by
 speckle methods. Weigelt \& Ebersberger (1986) and Hofmann \& Weigelt
 (1988) found 3 objects close to the central star (0.1-0.2''
 separation); first HST UV observations of the speckle objects were
 reported by Weigelt et al. (1995).  They appear very compact in
 far-red light; but since they are moving outward (Weigelt et al. 1995
 \& 1996) and have forbidden lines in their spectra (Davidson et
 al. 1995), they must be ejected clouds rather than companion stars.
 In H$\alpha$ the sub-arcsecond structure of $\eta$ Carinae is even
 more complex, showing an arm-shaped feature in the north (Weigelt et
 al. 1996).

Polarization observations have shown that $\eta$ Carinae is
intrinsically polarized (Visvanathan 1967, Marraco et
al. 1993). Warren-Smith et al. (1979) showed that the total
polarization of the Homunculus is always perpendicular to the direction to
the central object as expected for Mie scattering by dust grains. In
the outer regions the degree of polarization reaches up to 40\% while in
the inner regions it is well below 10\% (see also Meaburn et
al. 1993). Dust is expected to form at roughly the same distance
from the star as the speckle objects.

To extend those studies we have built a polarimeter for our speckle
camera and are now for the first time able to obtain high-resolution
polarimetric information at optical wavelengths with ground-based
telescopes. Here we report on results we obtained during a first
test-run of our speckle imaging polarimeter where we observed $\eta$
Carinae with an H$\alpha$ filter.

\section {Observations}
\subsection{The polarimeter}
Our new polarimeter consists mainly of a rotatable, achromatic
$\lambda/2$-retardation mica plate in front of a fixed polarization
filter. If rotated by an angle $\alpha$ the $\lambda/2$ plate rotates
the polarization vector of the incident light by $2\alpha$. The usable
wavelength range is 450-800 nm with an error for the retardation of
$2\%$ of $\lambda$/2 over the whole wavelength range and a
transmission of $\sim70\%$. The polarization filter has a transmission
of $\sim33\%$ and a polarization degree of $>99.99\%$ over a
wavelength range of 450-750nm. The $\lambda/2$-plate is mounted on a
remote-controlled rotator with a step motor of 1/500 Degree
resolution. Filter, rotator and $\lambda/2$-plate were installed on a
single mount that was inserted into the optical axis in front of the
telescope focus of our camera. The use of a fixed polarization filter
basically eliminates the effects of depolarization in the camera, as
the polarization vector of the light entering the camera has always
the same orientation. Circular polarization can not be measured.

\subsection{Observing strategy}
$\eta$ Carinae was observed with the ESO 2.2 m telescope in Chile on
March 12, 1995 between ST 12:20 and ST 12:50 with an improved version
of our MPIfR speckle camera (Baier \& Weigelt 1983) using a 30 nm wide
H$\alpha$ filter and 20 fold magnification giving us a field of view
of $\sim6$ arcsec. In total we took 4800 images of 50 ms exposure time
at a frame rate of 4 images/sec including flatfield and dark
images. The observation was split into 16 sections of 300 images
each. After each section we rotated the $\lambda/2$-plate by
$22.5^\circ$ corresponding to a rotation of $45^\circ$ of the incident
polarization vector. We finally obtained four independent measurements
of the polarized intensity for each of the four possible orientations
of the polarization vector ($0^\circ$, $45^\circ$, $90^\circ$, and
$135^\circ$). This has the benefit that we have a full rotation of the
$\lambda/2$-plate and two rotations of the polarization vector with
respect to the polarization filter which helps to detect and reduce
the effects of any rotational asymmetries in the polarimeter. A second
advantage of taking interleaved images is that the images for each
$45^\circ$ rotation of the polarization vector -- if added together --
are taken quasi-simultaneously thus a slow monotonic change in the
seeing conditions affects all images in a similar way. This is
especially important for the reconstruction of speckle
images. Finally, we want to note that the use of a speckle camera does
have another intrinsic advantage over conventional imaging
polarimetry: by evaluating the total intensities of all images we can
monitor the short-time variability of the atmospheric transmission.

After the observation we measured a flatfield to determine the
photonbias (Pehlemann et al. 1992) and a single star (SAO251486) to obtain
the speckle transfer function. The same observations as for $\eta$
Carinae were performed thereafter with a nearby cluster member (HDE
303308) for comparsion.

\begin{figure*}
\centerline{
\bildh{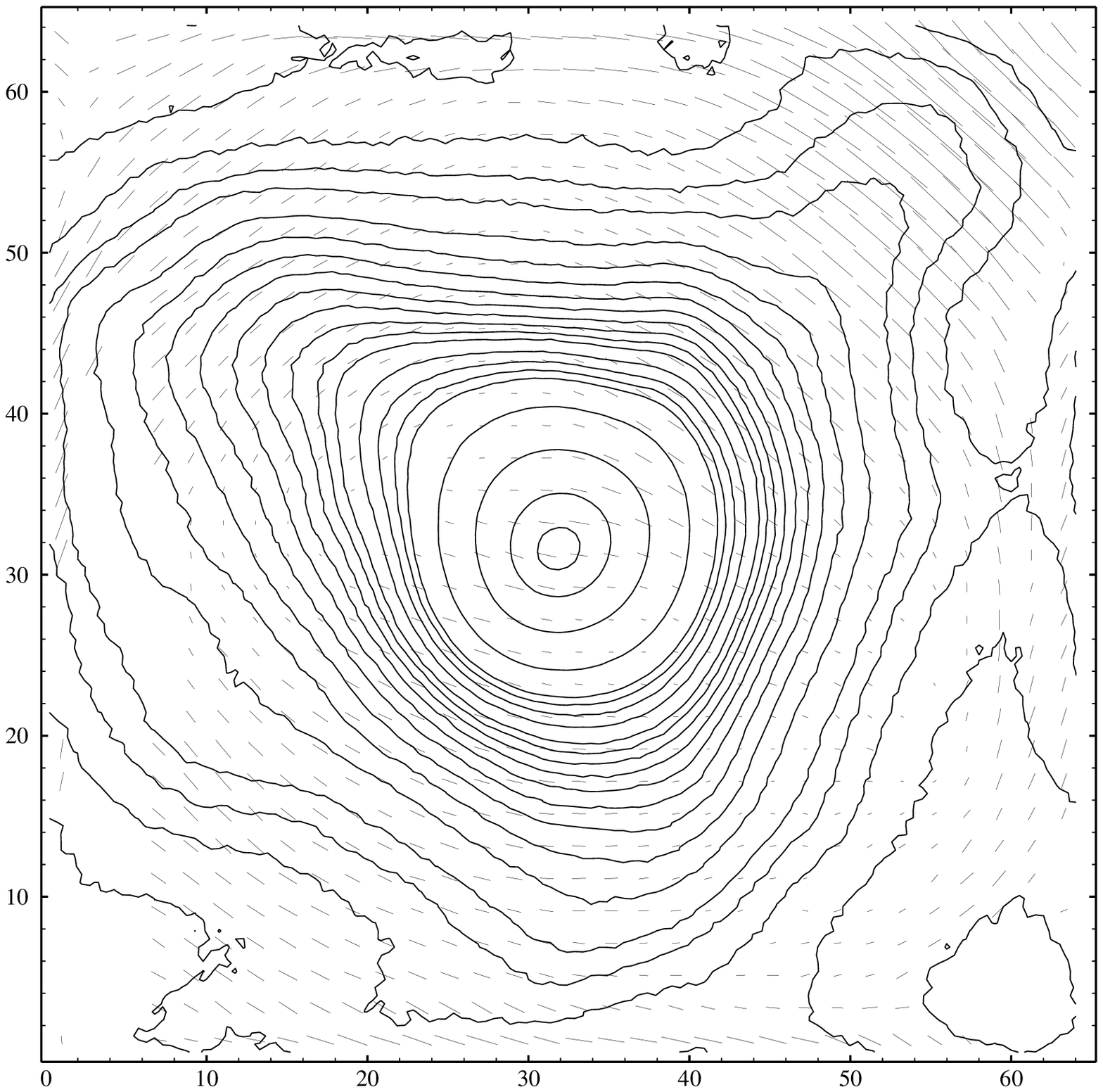}{5.5cm}{bbllx=2.3cm,bblly=5.4cm,bburx=19.2cm,bbury=22.1cm}
\bildh{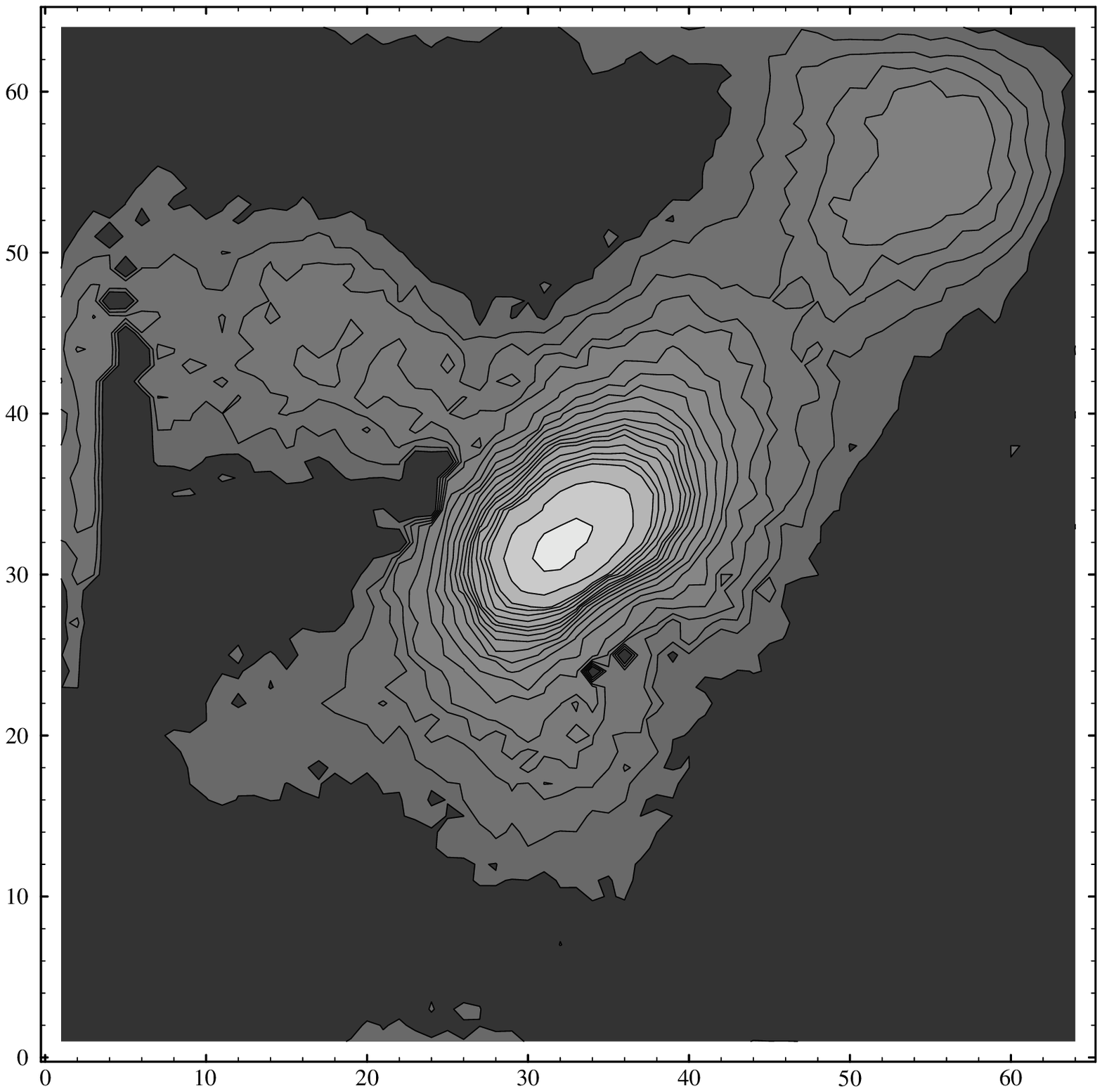}{5.5cm}{bbllx=2.3cm,bblly=5.5cm,bburx=19.2cm,bbury=22.2cm}
\bildh{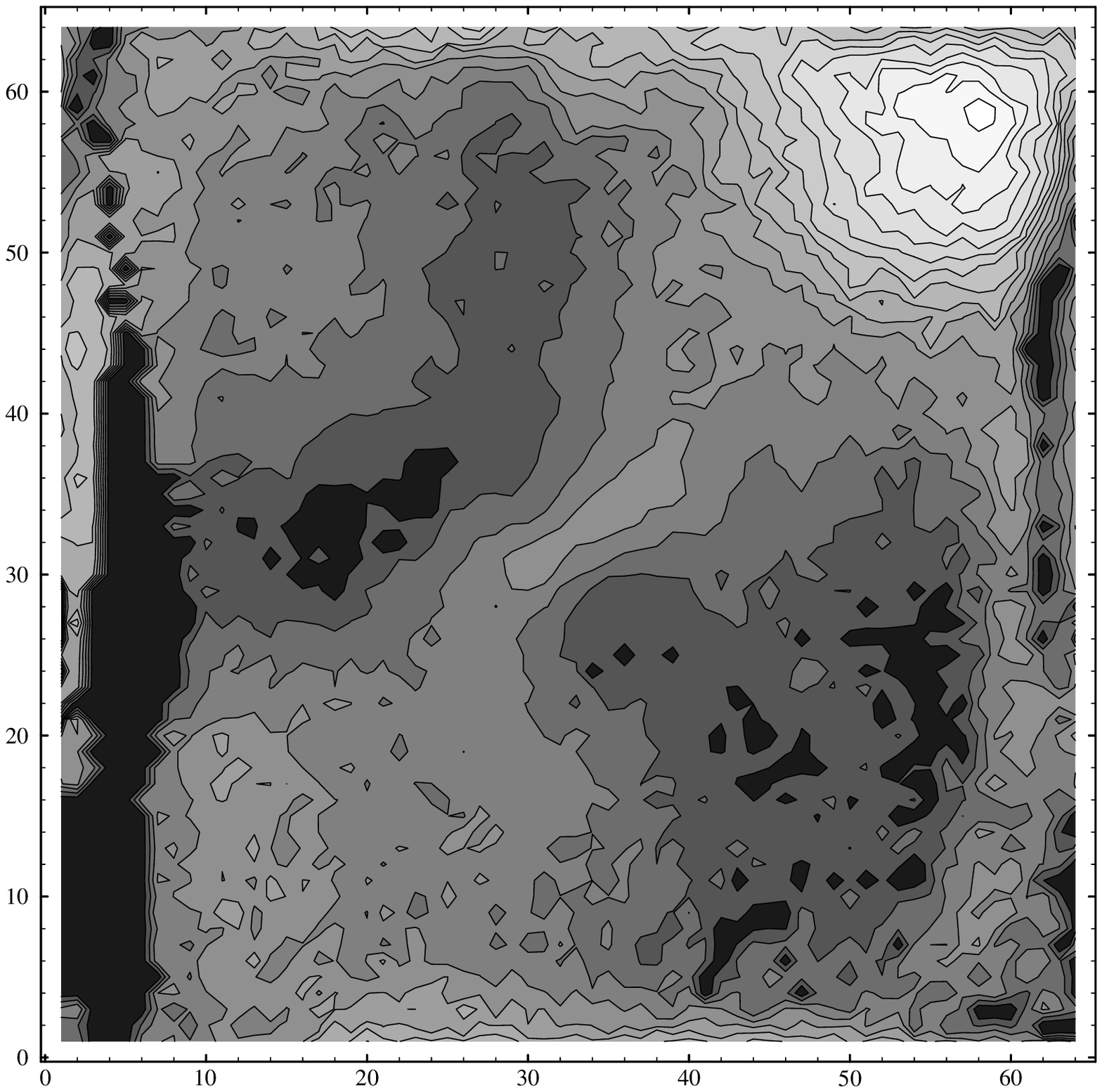}{5.5cm}{bbllx=2.3cm,bblly=5.5cm,bburx=19.2cm,bbury=22.2cm}
}
\caption[]{a) Crosscorrelation-centered $6\times6$ arcsec image of $\eta$
Carinae and its polarization -- north is up and east to the left. The
short lines indicate the orientation of the E-vector and the length is
proportional to the degree of polarization.  b) The total polarized
flux of the left image. c) The per-cent polarization of the left
image. The images were zoomed from our $512^2$ format down to a $64^2$
format. The distortions at the edges of the images are consequences of
the centering procedure. }
\end{figure*}

\section{Data reduction}
\subsection{Total polarization}\label{totalpol}
Prior to the data reduction we combined the four small data sets for
each polarization angle into a large data set containing 1200 images
each. During the speckle-masking reconstruction of the images the
information of the total intensity is lost and therefore one has to
re-scale the intensity according to the intensities of the co-added
long-exposure images. One way to achieve this would be to simply
co-add all images and determine the degree of polarization from the
four long-exposure images. However, using this method, we would loose
the valuable informaton of the short-term variability of our
data. Therefore, we extracted the (image-intensifier dark subtracted)
integrated intensities of each image and plotted them in the order
they had been obtained originally. Although the seeing was relatively
stable and we had perfect weather conditions, we found small variations
of the integrated short-exposure image intensities of $\eta$ Carinae
of the order of a few per cent, clearly exceeding the instrumental
noise in amplitude. The run of intensities can be described by a
constant upper envelope plus occasional, erratic dips. We interpret
this as short-term reduction of the atmospheric transmission from the
optimal value; an uneven distribution of those dips among the
different polarization angles would clearly affect any polarization
results based on the usual average (as in long exposures). A simple
way to eliminate the influence of those dips is to take the
median value of an upper constant envelope of the intensities,
e.g. the brightest $N$ of all images. Indeed we found that by
varying $N$ and calculating the deviation of the polarized intensity
from the expected Sinus shape the error has a well defined minimum
around $N=16\%$, yielding a degree of polarization of $P=4.06\%$ and a
position angle (PA) for the E-vector of $\theta_{\rm P}=73.5^\circ$
for our field of view. Nevertheless, the results differed only by
$\pm0.025\%$ in $P$ and $\pm0.5^\circ$ in $\theta_{\rm P}$ over a
range of $N=10\%-60\%$ (which would include the `normal'
median).\footnote{For this extended object, however, the total
polarization vector of our small square aperture can not easily be
compared with usual large round apertures.}

The situation for the cluster star HDE 303308 was slightly different
as it is substantially fainter than $\eta$ Carinae and the intensity
variations are mainly dominated by photon and instrumental noise.
Hence, we took the usual median intensities for the four polarization
angles at $N=50\%$ -- where the minimum error was found -- yielding
$P=2.76\%$ and $\theta_{\rm P}=102\%$.  Here, the systematic
variations by changing $N$ are $\pm0.3\%$ and $\pm5^\circ$
respectively. We noticed a 100 times fainter companion $3.4\arcsec$
away from HDE 303308 at PA $231^\circ$.

\subsection{Image reconstruction and polarization maps}
From the four combined data sets we reconstructed four images using
the basic speckle-masking technique as described in Weigelt (1977),
Lohmann et al. (1983), and Hofmann \& Weigelt (1986). To ensure that
the reconstructions of all images were done in an identical manner, we
used the automatized speckle processing package (ASP) recently
developed at the MPIfR (Falcke et al. 1996).

The basic scheme of the automated speckle data reduction process
consists of the following steps: image intensifier dark current and
flatfield corrections, detection of ion contamination, seeing
selection, calculation of crosscorrelation-centered images,
photon-bias compensation, bispectrum calculation, and image
reconstruction. Thus we obtained images at different resolutions up to
the diffraction limit. Our diffraction-limited H$\alpha$ total
intensity image confirms the detections by Weigelt et al. (1996)
and especially shows the northern $H\alpha$-arm and its
blobs and even the weak features to the SE and NE of the nucleus.

To obtain our polarization map, we first determined the position of
the central peak in each image by fitting a two-dimensional gaussian
and shifted each image onto a common center. We then normalized the
total intensity of the reconstructed images according to the values
found in Section \ref{totalpol}.  From the four shifted and normalized images
corresponding to the four polarization angles we determined the Stokes
parameters for each pixel and the standard deviation from a sinusoidal
distribution. In the polarization maps we left out all vectors having
errors $>30\%$ although mostly the errors are $<10\%$. For the vector
maps we usually also combined four pixels for one vector. Most
affected by errors are the polarization angles in the fainter parts of
the reconstructed image and we have not made any attempt to correct
the interstellar polarization in the H$\alpha$ line towards $\eta$
Carinae.

\section{Results}
In Fig. 1a we show the contours of the crosscorrelation-centered
$6\times6$ arcsec total intensity image of $\eta$ Carinae overlayed
with the vectormap of the polarization. The tangential pattern of the
polarization vectors is in good agreement with the results by
Warren-Smith et al. (1979); this is best seen in the NW spur where the
polarization reaches values of 20-40\%. The bipolar nature of the
Homunculus becomes apparent in the polarized light (Fig. 1b\&c) and
there is also a bar along the minor axis (PA 42$^\circ$) towards the NE,
which is already present in the total intensity map but becomes a
marked feature in the polarized intensity map (Fig. 1b). There may
also be a very weak SW counterpart which is, however, not well
visible in the contour plots. Figure 1c shows that the degree of
polarization is asymmetric with respect to the center and is lower
towards the SE and higher in the NW. The degree of polarization seems
to be reduced in a strip along the minor axis and the central object
appears elongated. Such a pattern can be found if the central star
itself is polarized (Els\"asser \& Staude 1978; Gledhill 1990). 

In Fig. 2a we show the contour map of the H$\alpha$ speckle-masking
reconstruction of the inner arcsecond of the Homunculus and its
polarization. To increase the SNR of the polarization map we have not
reconstructed the image up to the diffraction limit but with a lower
resolution of 0.11 arcsec. At the central peak the polarization is
$P=9.1\%$ and $\theta_{\rm P}\simeq80^\circ$. It is remarkable that
the largest part of the H$\alpha$ arm and the four speckle objects A-D
are in a region of relatively low polarization around $10\%$. The PA
of the polarization vectors of the three speckle objects B-D are similar
to those of the central star and almost perpendicular to the radial
axis towards the center. The polarization increases strongly towards
the NW where it still is perpendicular to the radial axis and
$P=20-40\%$.  This feature connects well into the NW spur seen already
in Fig.1a.

It is also noteworthy that the total intensity of $\eta$ Carinae in
the high-resolution image is sharply reduced below the minor (NE-SW)
axis.  In the polarized intensity map (Fig. 1b) we do see several
co-linear blobs along the same axis which are symmetric around the
center.  We note that this linear feature also connects the NE end of
the northern arm and the central star. In a larger field of view it
also continues smoothly into the bar noted already in Fig.1 while
there is no such strong feature in the SW on the larger scale.

\section{Summary and Discussion}
\begin{figure*}
\centerline{
\bildh{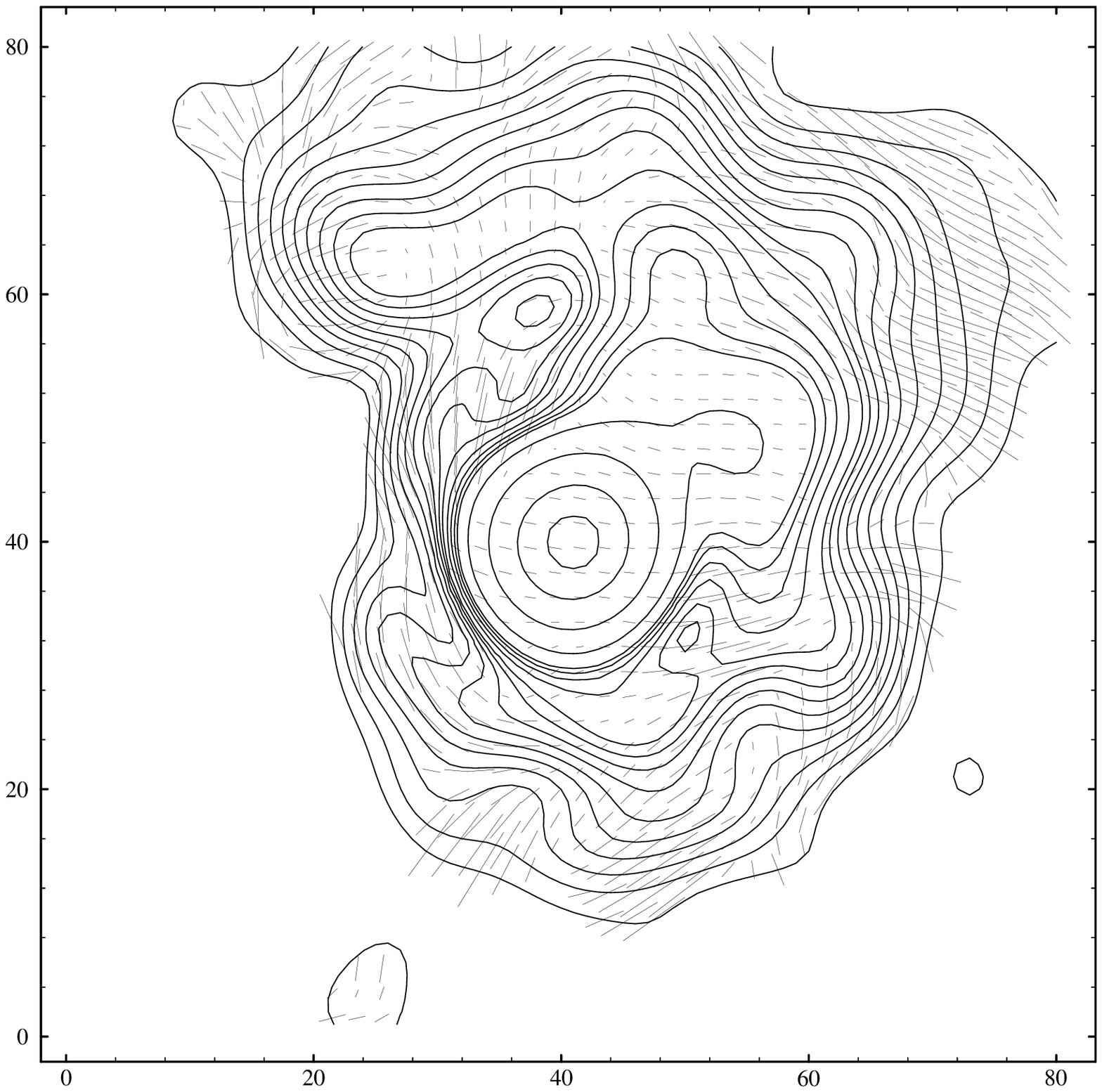}{8cm}{bbllx=2.3cm,bblly=5.5cm,bburx=19.2cm,bbury=22.2cm}
\bildh{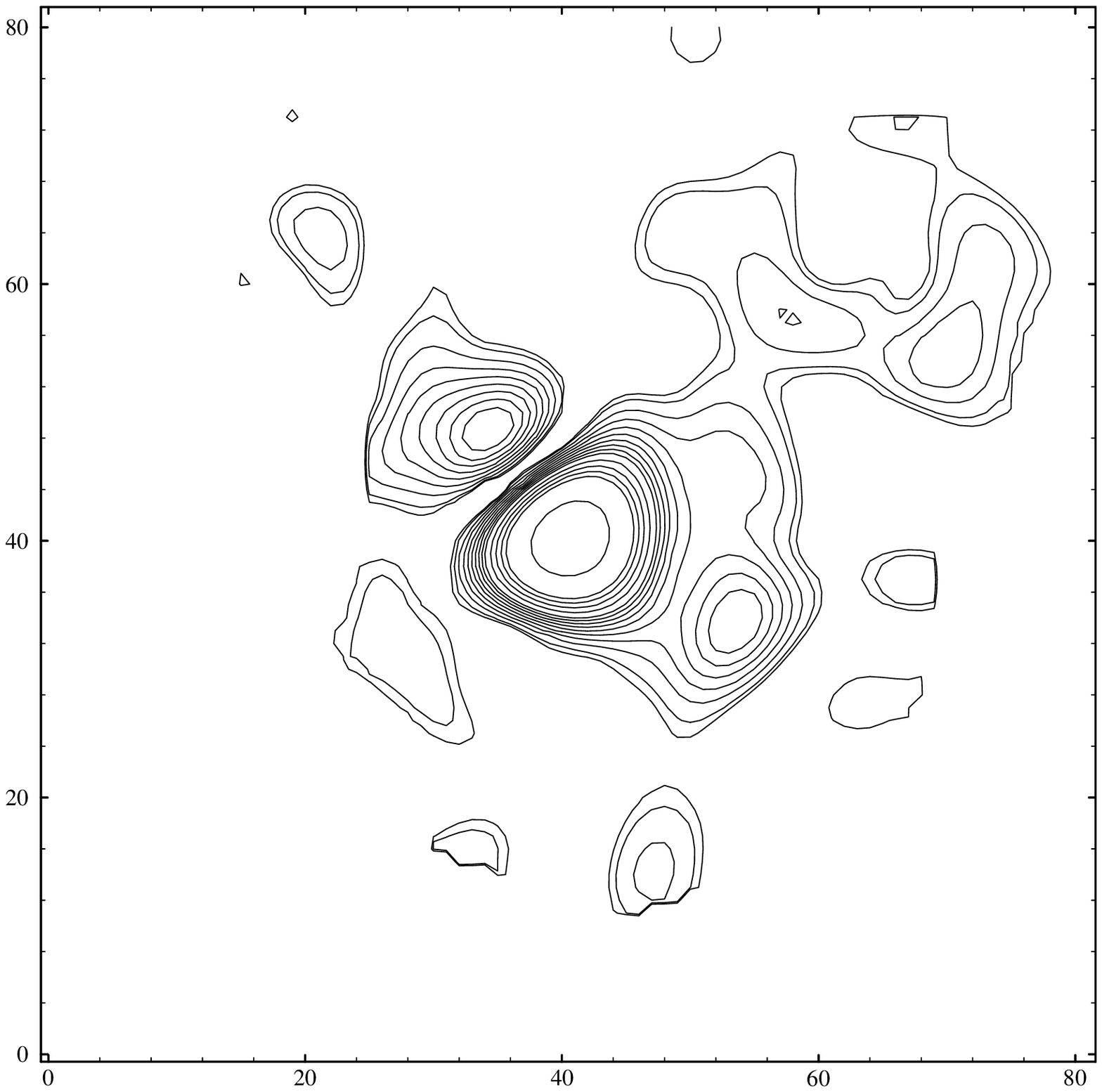}{8cm}{bbllx=2.3cm,bblly=5.5cm,bburx=19.2cm,bbury=22.2cm}
}
\caption[]{a) Contour map of the H$\alpha$ speckle-masking
reconstruction and E-vector map of $\eta$ Carinae for a
$0.94\times0.94$ arcsec field (north is up, east to the left). The
resolution is artificially degraded to $60\%$ of the diffraction limit
corresponding to $0.11$ arcsec for the benefit of a higher SNR.  The
short lines indicate the orientation of the E-vector and the length is
proportional to the degree of polarization.  b) The total polarized
flux of the left image. The hole at (65,65) in the NW structure corresponds to a
high-error pixel which was left out.}
\end{figure*}

H$\alpha$ observations of the LBV $\eta$ Carinae during a first
test run of our polarimeter have demonstrated the feasibility of
optical speckle imaging polarimetry.  Degree and PA of the polarization of the
reference star HDE 303308 agree within the errors with the literature
values (Visvanathan 1976; Marraco et al. 1993), the outer polarization
pattern of our $\eta$ Carinae image itself match well the inner
structure found by Warren-Smith et al. (1979) and the high-resolution
reconstruction of $\eta$ Carinae confirms the structures
detected by Weigelt et al. (1996). Several large-scale features
continue down to the small scales.

Along the minor axis of the Homunculus we have detected a bar to the
NE of $\eta$ Carinae in the polarized light and we find a linear
symmetric structure in the inner arcsecond which is oriented in the
same direction. This, together with the sharp intensity drop from the
NW to the SE in the high-resolution image and the constriction of the
central contours on the larger scale may be indicative of the presence
of a dusty equatorial disk around $\eta$ Carinae with its rotation
axis along the major axis of the Homunculus. The NE bar and the
central arcsecond bar could then be interpreted as scattered light
from the surface of the disk. This may explain the PA and the high
degree of the polarization in the NE and SW blobs in Fig. 2b.  The
absence of a SW counterpart to the NE bar on the large scale (while
present at the sub-arcsecond scale) might be explained by a warped or distorted
geometry. The up-turn of the NE end of the H$\alpha$ arm seen in this
paper and by Weigelt et al. (1996) could be indicative of a physical
connection between the arm and the putative disk.

The usual explanation of the brightness contrast between the parts
above and below the minor axis would be obscuration by the disk.
The SE side of the homunculus polar axis points obliquely
     {\em towards} us, as seen in velocity data (Thackeray 1961, Meaburn
     et al. 1987) and in modern high-resolution images (Duschl et 
     al. 1995, Humphreys \& Davidson 1994).  Material oriented
     along the polar axis would therefore be more visible
     on the SE (nearer) side and obscured on the NW side, exactly the
     opposite to what is observed near the star.  Hence we 
     conclude that most of the small-scale structure discussed
     above is essentially equatorial rather than polar:  we are
     seeing the most visible inner NW parts of the equatorial 
     ejecta-disk pointing towards us. 

     This picture is consistent with the low polarization of the
     speckle objects and the H-alpha arm ($\la$ 10\%), since Mie
     scattering produces lower polarization at relatively small
     scattering angles.  We also point out that the polarization
     properties of the central star and the speckle objects B-D are
     basically indistinguishable even though the latter are clearly
     scattered light from the nucleus (Davidson et al. 1995), hence in
     the nucleus we may see in the polarized light a dusty envelope
     (or disk) rather than the naked star alone, and the speckle
     objects B-D as well as the H$\alpha$ arm might well be part of
     the disk and its radial streamers (see Duschl et al. 1995). That
     the central star is most likely obscured was already evident from
     earlier observations where it was shown that the bright nucleus
     is extended and much too faint with respect to the surrounding
     speckle objects B-D (Weigelt et al. 1995).

Some of the H-alpha light observed in the speckle objects and the
     H-alpha arm may be emitted there, rather than scattered -- unlike
     the case for larger size scales in the homunculus.  This would of
     course be consistent with low polarizations.  Further
     observations in continuum light may show larger amounts of
     polarization, since most {\em continuum} light in the blobs is
     expected to be scattered from the central star. In addition to
     the speckle objects and the H$\alpha$ arm we also find at the
     sub-arcsecond scale something which might be the continuation of
     the NW spur seen in Figure 1. It appears in the polarized
     intensity image (Fig. 2b) but is barely visible in normal
     intensity (Fig. 2a).  The polarization of this feature is very
     high (up to 40\%), indicating large scattering angles and
     suggesting that this feature is part of the low-density, highly
     polarized NW part of the Homunculus nebula rather than a part of
     the disk.

We conclude that our findings support the model of an equatorial disk
surrounding $\eta$ Carinae. The presence of such a disk can be very
important for the angular momentum loss in the late phases of massive
stars -- if it is an excretion disk, as indicated by the radial
streamers (often referred to as jets). Still, we have to await further
observations as from this test run we had only a limited number of
images available for our reconstruction. 

\acknowledgements HF was supported by a Max-Planck Stipend. We thank
P.L.~Biermann for stimulating comments and R.~\"Osterreicher and
C.~M\"ollenhoff for helpful discussions on imaging polarimetry.

\end{document}